\documentclass[review,3p,times]{elsarticle}

\usepackage{amssymb}
\usepackage{amsmath}
\usepackage{lineno}
\usepackage{booktabs}
\usepackage{longtable}
\usepackage{siunitx}
\usepackage{graphicx}
\usepackage{subcaption}
\usepackage{float}
\biboptions{sort&compress}

\journal{Nuclear Instruments and Methods in Physics Research Section A: Accelerators, Spectrometers, Detectors and Associated Equipment}

\begin{document}

\begin{frontmatter}

\title{Development and Experimental Characterization of a Compact $^{228}$Th--D$_2$O Photoneutron Source for Low-Energy Nuclear Recoil Calibration}

\author[inst1]{Yueqiang Tian}

\author[inst1]{Chen Cheng\corref{cor1}}
\ead{chengch27@buaa.edu.cn}

\author[inst1]{Yuxin Cui}
\author[inst1]{Xunan Guo}
\author[inst1]{Zichao Guo}
\author[inst1]{Yao Tong}
\author[inst2]{Pingquan Wang}
\author[inst1]{Hailun Yuan}
\author[inst2]{Hui Zhang}

\author[inst1]{Xiaopeng Zhou\corref{cor2}}
\ead{xpzhou@buaa.edu.cn}

\affiliation[inst1]{organization={School of Physics, Beihang University},
            addressline={37 Xueyuan Road},
            city={Beijing},
            postcode={100191},
            country={China}}

\affiliation[inst2]{organization={National Institute of Metrology},
            addressline={No. 18, North Third Ring Road East},
            city={Beijing},
            postcode={100029},
            country={China}}

\cortext[cor1]{Corresponding author}
\cortext[cor2]{Corresponding author}

\begin{abstract}
Calibrating the few-keV nuclear-recoil response of liquid xenon detectors calls for a low-energy neutron source. We have developed and characterized a $^{228}$Th--D$_2$O photoneutron source that converts the 2.615~MeV $\gamma$ rays from $^{208}$Tl into sub-MeV neutrons through the D($\gamma$,n)p reaction. After optimizing the D$_2$O geometry to balance neutron production against escape efficiency, we validated the chosen configuration experimentally. The forward emission rate is $1734 \pm 76~(\mathrm{stat.}) \pm 65~(\mathrm{sys.})~\mathrm{h}^{-1}$, and the spectrum unfolded from Bonner sphere data is concentrated at $0.17$--$0.22~\mathrm{MeV}$. In a 220~h exposure of PandaX-20T, the same configuration should provide about 1000 selected nuclear-recoil events. These results support the use of a long-lived $^{228}$Th--D$_2$O source for low-energy neutron calibration of large liquid xenon detectors.
\end{abstract}

\begin{highlights}

\item A compact $^{228}$Th--D$_2$O photoneutron source was designed and characterized for ultralow nuclear recoil calibration.


\item Bonner sphere spectrometry: 0.17--0.22~MeV peak, all neutrons below 1~MeV.



\item Calibrating for neutrinoless double-beta decay ($0\nu\beta\beta$) region of interest electron recoil at the same time.

\end{highlights}


\end{frontmatter}

\section{Introduction}
\label{sec:introduction}

Astrophysical and cosmological observations indicate that dark matter constitutes approximately 85\% of the matter density of the Universe. Among various dark matter candidates, weakly interacting massive particles (WIMPs) remain one of the primary targets of direct detection experiments. However, coherent elastic neutrino--nucleus scattering (CE$\nu$NS) induced by solar, atmospheric, and diffuse supernova neutrinos represents an irreducible background for future direct dark matter searches, ultimately limiting their sensitivity, commonly referred to as the neutrino floor \cite{Freedman1974,Billard2014NeutrinoFloor,OHare2021}. Liquid xenon time projection chambers (LXe TPCs) have become one of the leading technologies for direct dark matter searches due to their large target mass, excellent self-shielding capability, and detection of prompt scintillation photons (S1) and ionization electrons (S2) \cite{XENONnT2024}. The ratio between S2 and S1 provides powerful discrimination between electron recoils (ER) and nuclear recoils (NR), enabling sensitive searches for rare interactions.

The same CE$\nu$NS process that constitutes an irreducible background for dark matter searches is also an important physics signal in its own right. In addition to dark matter searches, LXe TPCs have therefore demonstrated the capability for low-energy neutrino measurements through CE$\nu$NS detection. XENONnT, PandaX-4T, and LUX-ZEPLIN (LZ) experiments have reported evidence for solar ${}^{8}$B neutrinos through CE$\nu$NS in LXe~\cite{XENON8B2024,PandaX8B2024,LZ:2025igz}. PandaX-4T has also been used to search for other low-energy dark matter signatures, including solar-boosted dark matter through ER signals~\cite{Shen2025SolarBoostedDM}. These measurements highlight the importance of accurately understanding the low-energy response of large LXe detectors, particularly in the few-keV energy region where different rare-event signals can overlap.

Accurate low-energy calibration is therefore an essential component of the physics program of large LXe experiments. Low-energy detector response, signal reconstruction, event selection, and ER calibration have been studied using dedicated radioactive sources in several LXe experiments. For example, internal radioactive-source calibration has been demonstrated in the PandaX-II experiment using ${}^{220}$Rn and ${}^{222}$Rn gaseous sources, enabling detailed studies of low-energy ER responses and radioactive backgrounds~\cite{Ma2020PandaXRnCalibration}. The XENON1T experiment also demonstrated the use of an internal ${}^{37}$Ar source, providing monoenergetic energy depositions at 2.82~keV and 0.27~keV and enabling characterization of the detector response close to the detection threshold~\cite{XENON1TAr37}. An ${}^{37}$Ar source has also recently been developed and characterized for LXe detector calibration~\cite{Guo2026Ar}. These developments illustrate the effectiveness of dedicated radioactive sources for establishing the low-energy ER response of LXe detectors.

In contrast, precise calibration of low-energy NRs remains considerably more challenging. Neutron calibration has been extensively employed by LXe experiments to characterize the scintillation and ionization response of NR~\cite{Akerib2016DD,Aprile2018NR}. The PandaX-II experiment performed a systematic measurement of the LXe response to low-energy ERs and NRs using radioactive calibration sources, covering the energy ranges relevant to low-energy detector response studies~\cite{Yan2021PandaXCalibration}. The LUX experiment has also demonstrated NR calibration at sub-keV energies, highlighting the increasing importance of precise NR response measurements near the detector threshold~\cite{Akerib2025LUX}. At the same time, the stringent background requirements of large LXe detectors motivate detailed characterization and mitigation of low-energy backgrounds. In particular, neutron-induced NRs constitute an important background component in LXe dark matter experiments and have been studied in detail for PandaX-4T~\cite{Huang2022PandaXNeutronBackground}. Similar background considerations have also been addressed by the LZ experiment~\cite{LZ2023Background}.

Conventional neutron sources such as ${}^{252}$Cf, AmBe, and deuterium--deuterium (D--D) neutron generators generally produce broad or relatively high-energy neutron spectra. The resulting recoil-energy distributions span a wide range, making it difficult to precisely determine the LXe response in the few-$\mathrm{keV}_{\mathrm{nr}}$ region (where the subscript ``nr'' denotes NR energy)~\cite{Mannhart1987,Marsh1995AmBe,Liskien1973DD}. Monoenergetic or quasi-monoenergetic neutron sources with neutron energies below 1~MeV are therefore highly desirable for precise low-energy NR calibration.

Photoneutron sources provide an effective approach for producing quasi-monoenergetic neutrons by utilizing photons with energies close to the reaction threshold. In particular, ${}^{88}$Y--Be photoneutron sources have been developed and successfully applied for low-energy NR calibration in LXe experiments~\cite{Collar2013,XENONnT88YBe2026,LZYBe2025}. The recent XENONnT measurement used a 152~keV photoneutron source to characterize the LXe light and charge response from 0.3 (0.7) to 5.0~$\mathrm{keV}_{\mathrm{nr}}$, demonstrating the value of photoneutron calibration for the low-energy physics program of large LXe detectors~\cite{XENONnT88YBe2026}. The LZ collaboration has likewise developed a photoneutron calibration program for low-energy NR response measurements~\cite{LZYBe2025}. However, the relatively short half-life of ${}^{88}$Y limits the long-term operational lifetime of such sources, motivating the development of alternative photoneutron sources based on longer-lived radionuclides.

The $^{228}\mathrm{Th}$ source provides an alternative approach by utilizing the intense 2.615~MeV $\gamma$ ray emitted from $^{208}\mathrm{Tl}$. The photon energy exceeds the deuteron photodisintegration threshold, allowing neutron production through the $\mathrm{D}(\gamma,\mathrm{n})\mathrm{p}$ reaction, which has a $Q$ value of $-2.224$~MeV. The absolute intensity of the 2.615~MeV $\gamma$ emission and the photodisintegration cross section of deuterium have been measured previously \cite{Martin2007ENSDF208Tl,BIPM5Vol7,Marin1953}.

Compared with ${}^{88}$Y-based photoneutron sources, ${}^{228}$Th provides a substantially longer half-life, improving source lifetime and operational stability \cite{Chechev2011,BIPM5Vol7}. Moreover, the 2.615~MeV $\gamma$~ray itself provides an additional calibration capability for LXe detectors. Its energy is close to the $Q$ value of the neutrinoless double-beta decay ($0\nu\beta\beta$) of ${}^{136}$Xe ($Q_{\beta\beta}=2.458$~MeV) \cite{Redshaw2007}, making it a valuable benchmark for calibrating the high-energy ER response. Therefore, a ${}^{228}$Th--$\mathrm{D_2O}$ source can potentially provide both low-energy NR calibration via photoneutron production and high-energy $\gamma$-ray calibration for LXe detector response.

In this work, we report the development and characterization of a compact ${}^{228}$Th--D$_2$O photoneutron source designed for low-energy NR calibration of LXe detectors or other similar detectors. The source geometry was optimized using Geant4 simulations with the Low Energy Nuclear Data (LEND) physics list~\cite{Agostinelli2003,Allison2016}. The forward neutron emission rate and energy spectrum were experimentally measured using a ${}^{3}$He proportional counter and a Bonner sphere spectrometer (BSS), respectively. The measured source performance was further evaluated through Monte Carlo simulation of the PandaX-20T detector calibration~\cite{PandaXxT2025}.

\section{$^{228}$Th--D$_2$O Photoneutron Source}
\label{subsec:th228_source}
The $^{228}$Th--D$_2$O photoneutron source operates based on the deuteron photodisintegration reaction. In the decay chain of $^{228}$Th ($T_{1/2}\approx1.9$~years), the daughter nuclide $^{208}$Tl emits an intense 2.615~MeV $\gamma$ ray \cite{Martin2007ENSDF208Tl,BIPM5Vol7,Chechev2011}. When these photons interact with deuterium nuclei in heavy water (D$_2$O), neutrons are produced through the reaction shown in Eq.~\eqref{eq:photo_disintegration}:

\begin{equation}
\gamma \; (E_\gamma = 2.615\;\mathrm{MeV})
+ \mathrm{D}
\rightarrow p+n,
\qquad Q=-2.224\;\mathrm{MeV}.
\label{eq:photo_disintegration}
\end{equation}

The reaction threshold is determined by the deuteron binding energy of approximately 2.224~MeV, leaving an available kinetic energy of about 0.391~MeV for the outgoing nucleon pair \cite{Marin1953}.

\begin{figure}[htbp]
\centering
\includegraphics[width=0.70\linewidth]{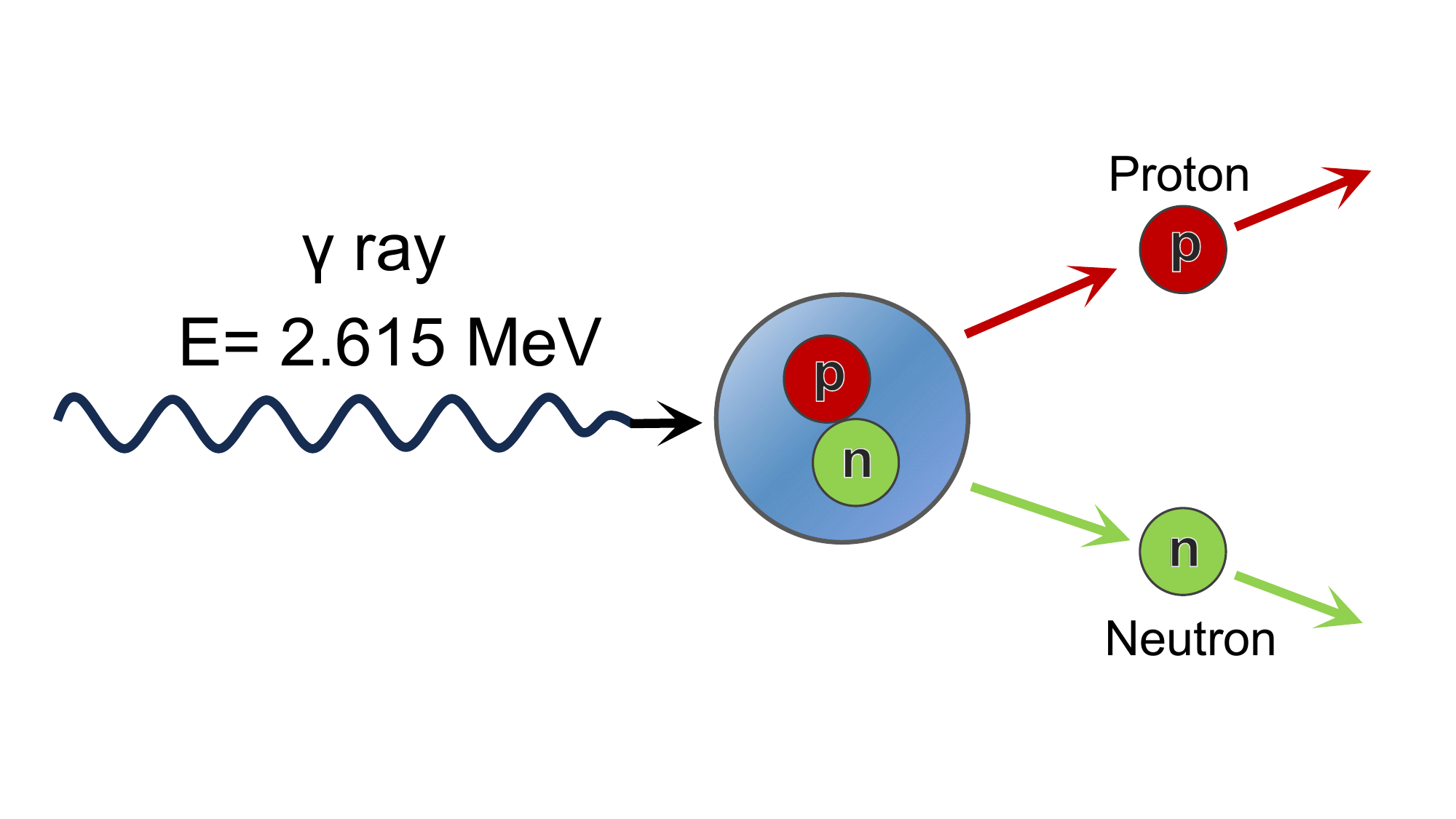}
\caption{Schematic illustration of the photodisintegration process. A 2.615~MeV $\gamma$ ray from $^{208}$Tl in the $^{228}$Th decay chain interacts with a deuteron and produces a proton and a neutron. The reaction threshold is determined by the deuteron binding energy of 2.224~MeV, leaving an excess kinetic energy of 0.391~MeV shared between the outgoing proton and neutron~\cite{Marin1953}.}
\label{fig:PhotoNuclear}
\end{figure}

Neutron production is governed by direct two-body photodisintegration rather than a compound-nucleus or statistical evaporation mechanism \cite{Marin1953}: a single photon directly breaks the weakly bound deuteron into a proton and a neutron (Fig.~\ref{fig:PhotoNuclear}). Competing electromagnetic interactions of the 2.615~MeV $\gamma$-rays in $\mathrm{D_2O}$, such as Compton scattering off the atomic electrons of D and O, do not produce neutrons through this channel.

Because the incident photon energy is only 0.391~MeV above the deuteron binding energy, the available kinetic energy is tightly constrained by two-body kinematics. In the center-of-mass frame, the neutron and proton have equal and opposite momenta, and the limited kinetic energy is distributed between them according to energy and momentum conservation. Consequently, the emitted neutron energy remains in the sub-MeV range. The resulting neutron energy spectrum, including the effects of reaction kinematics and angular dependence, is evaluated using the Geant4 simulation described in Sect.~\ref{sec:energy_measurement}.

Unlike multistep reactions involving excited compound nuclei, this direct two-body photodisintegration process does not generate an excited residual nucleus undergoing subsequent particle evaporation or complex de-excitation cascades. As a result, the neutron emission is not accompanied by a broad high-energy neutron continuum or by significant additional high-energy $\gamma$ rays from nuclear de-excitation. More importantly, because the reaction is a two-body process with tightly constrained kinematics, the neutron energy is determined by the incident photon energy, the deuteron binding energy, and the emission angle. At a fixed detection angle, this yields a quasi-monoenergetic sub-MeV neutron population, making the $^{228}$Th--D$_2$O source particularly suitable for low-energy NR calibration.

It should be noted that the photodisintegration cross section of deuterium at 2.615~MeV is relatively small, approximately 1.38~mb \cite{Marin1953}. Therefore, the source is designed to provide a stable and well-characterized neutron yield rather than an extremely high neutron flux. For large LXe detectors, where precise response characterization is more important than achieving very high calibration rates, such a neutron yield is sufficient for dedicated calibration campaigns.

In addition to neutron production, the 2.615~MeV $\gamma$~ray emitted from $^{208}$Tl provides an independent calibration channel for LXe detectors. Its energy lies close to the $0\nu\beta\beta$ $Q$ value of $^{136}$Xe ($Q_{\beta\beta}=2.458$~MeV) \cite{Redshaw2007}, making it a valuable benchmark to calibrate the high-energy response of ER and the linearity of energy near the $0\nu\beta\beta$ region of interest \cite{Aalbers2024LZCalibration}. Consequently, the $^{228}$Th--$\mathrm{D_2O}$ source provides two complementary calibration capabilities: low-energy NR calibration through photoneutron production and high-energy ER calibration through the accompanying $\gamma$~emission. These two calibration applications do not necessarily require the same source deployment configuration. In particular, when the source is used for the calibration of NR, the accompanying $\gamma$ rays are treated as a background component, whereas the $^{228}$Th source itself can be deployed separately for dedicated calibration of $\gamma$ rays.

However, the intense 2.615~MeV $\gamma$~ray flux accompanying photoneutron production can introduce a substantial ER background during \textit{in situ} NR calibration. Therefore, additional shielding is required for NR calibration to attenuate the direct $\gamma$-ray flux while maintaining sufficient neutron transmission, as discussed in Sect.~\ref{sec:pandax_sim}. The optimized shielding suppresses the $\gamma$-induced ER background but does not completely eliminate the $\gamma$ component.

\subsection{Simulation and Optimization}
\label{subsec:simulation}

In the design of a $^{228}$Th--D$_2$O photoneutron source, D$_2$O serves simultaneously as neutron-production target and as a medium for neutron scattering and attenuation. This introduces a trade-off between neutron production and neutron escape efficiency, making the source geometry critical for optimizing the neutron output. Therefore, a dedicated Monte Carlo study was performed to optimize the geometry of D$_2$O and the position of the source.

The simulations were performed using Geant4 with the LEND physics list \cite{Agostinelli2003,Allison2016}. The geometry consisted of a horizontal cylindrical vessel made of high-purity quartz (SiO$_2$) with a diameter of 8~cm and filled with D$_2$O. The $^{228}$Th source was modeled as a disk with a radius of 1.25~cm. The radioactive decay of the source was simulated with the corresponding $\gamma$-ray emissions from the decay chain, with particular emphasis on the 2.615~MeV $\gamma$ line from $^{208}$Tl.

\begin{figure}[htbp]
\centering
\includegraphics[width=0.50\linewidth]{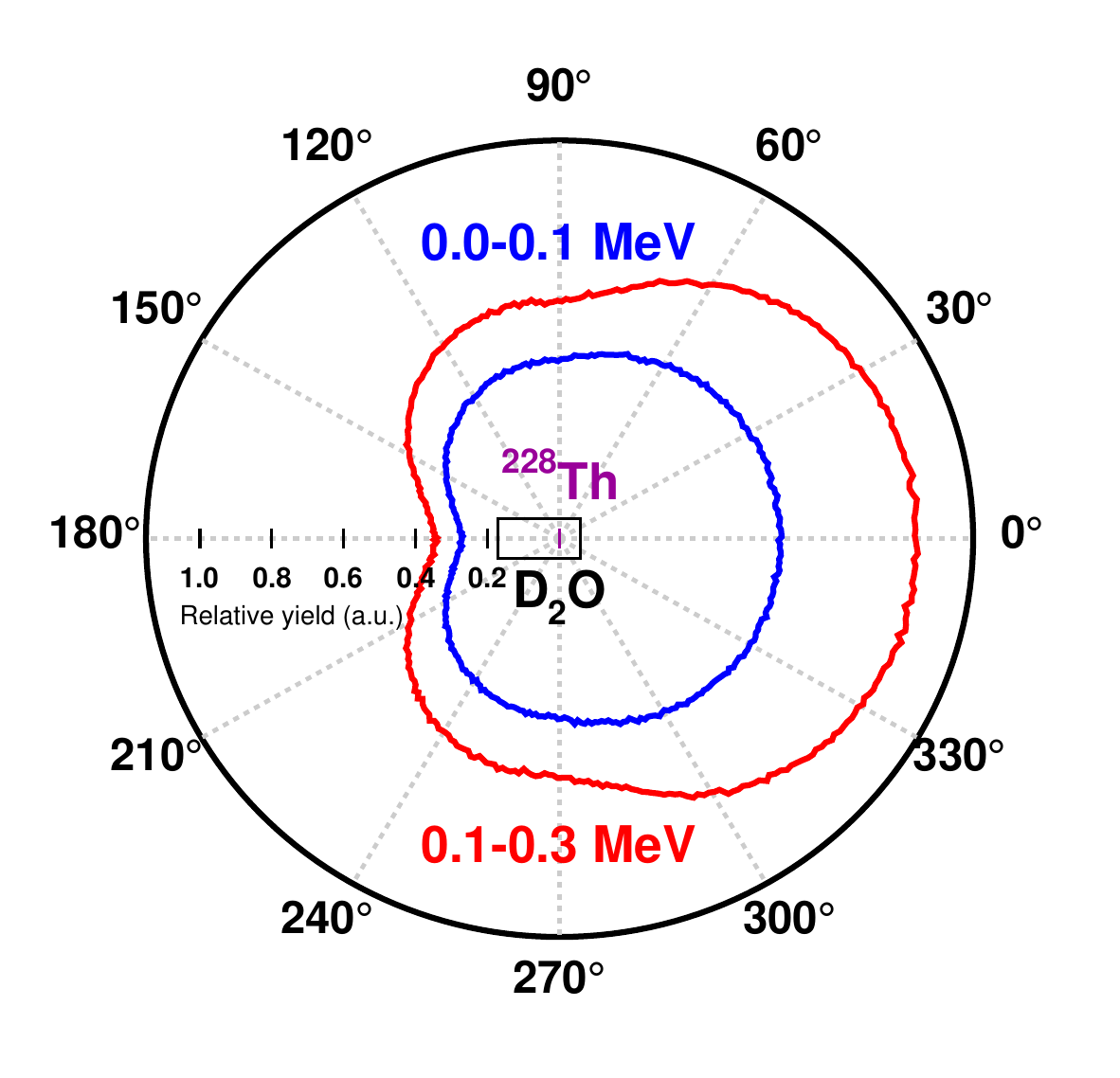}
\caption{Azimuthal distribution of escaping neutrons, plotted as a rose diagram of $\varphi_{XZ}=\mathrm{atan2}(P_x,P_z)$, with $0^\circ$ towards the detector. The radial coordinate is the relative neutron yield per $1^\circ$ azimuthal bin, normalized to the largest bin of the two groups so that their relative intensities can be compared; ticks along the $180^\circ$ axis are labelled 0.2--1.0. Because the simulated cross sections are modified with respect to the physical ones, only the relative angular shape is meaningful. A $^{228}$Th source sits inside the D$_2$O cylinder, 1~cm from the detector face. The forward enhancement near $0^\circ$ reflects shorter escape paths and reduced scattering. Red (0.1--0.3~MeV) corresponds to less-scattered neutrons consistent with the 0.17--0.22~MeV photoneutron peak, whereas blue (0--0.1~MeV) corresponds to multiply scattered neutrons. All escaping neutrons are below 1~MeV, satisfying the sub-MeV calibration requirement.}
\label{fig:rose_neutron}
\end{figure}

The optimization was performed by first fixing the source position and scanning the length of the D$_2$O cylinder to determine the minimum volume required for forward neutron emission to reach saturation; then, with this length fixed, the source position was varied along the longitudinal axis to find the location that maximizes forward emission. In the simulation, the end of the D$_2$O cylinder closer to the detector is defined as the ``forward'' direction, and the opposite end as the ``backward'' direction. Forward-emitted neutrons are defined as those that escape from the forward end face toward the detector. The results show that forward emission is maximized when the $^{228}$Th source is placed 1~cm from the forward end face. The general angular distribution of the escaping neutrons is shown in Fig.~\ref{fig:rose_neutron}.

The optimized source position results from the balance between neutron production and neutron escape probability. The 2.615~MeV $\gamma$ rays from the $^{228}$Th decay chain are emitted in all directions, and therefore photodisintegration reactions can occur throughout the D$_2$O volume rather than only near the forward surface. Since D$_2$O acts as both the neutron-production target and the neutron transport medium, neutrons generated near the forward end face have a higher probability of escaping directly towards the detector. Meanwhile, neutrons produced deeper inside the D$_2$O volume may either escape directly depending on their initial momentum direction or undergo elastic scattering before reaching the forward surface. The final escaping neutron distribution is therefore determined by both the initial neutron emission direction and the subsequent transport process in D$_2$O. The detailed energy-dependent effects of neutron transport and scattering are further discussed together with the measured neutron spectrum in Sect.~\ref{sec:energy_measurement}.

\subsection{Source fabrication}
\label{subsec:fabrication}

The $^{228}$Th source was fabricated by electrodeposition onto a stainless-steel disk with a radius of 1.25~cm and a thickness of 1~mm. Its activity was 72~kBq in July 2025 and its half-life is approximately 1.91~years \cite{Chechev2011,BIPM5Vol7}. The quoted activity refers to the $^{228}$Th parent activity; the activity of $^{208}$Tl is reduced by the branching ratio of the decay chain. In particular, with a $^{208}$Tl branching ratio of approximately 35.9\%~\cite{Martin2007ENSDF208Tl,BIPM5Vol7}, the corresponding $^{208}$Tl activity is about 25.9~kBq. The $^{228}$Th decay chain proceeds through multiple $\alpha$ and $\beta$ decays to stable $^{208}$Pb. Under secular equilibrium, the activity of each daughter nuclide is equal to that of its parent only when the daughter is produced with unit branching probability. In PandaX experiments, $^{228}$Th-based radioactive sources have also been used to introduce $^{220}$Rn into the LXe detector, allowing studies of the subsequent $^{216}$Po decay and its parent-daughter correlations \cite{Li2026Po216PandaX}.

During the decay chain, abundant $\alpha$ particles are emitted. To prevent these $\alpha$ particles from escaping the source and interacting with D$_2$O to induce unwanted $(\alpha,n)$ reactions, the radioactive material was hermetically encapsulated. Such $(\alpha,n)$ reactions are undesirable because they would introduce an additional neutron-production channel that is not part of the intended $^{2}$H$(\gamma,n)$p process and could therefore distort the measured neutron yield and energy spectrum. This encapsulation isolates the source from the surrounding D$_2$O and effectively stops the emitted $\alpha$ particles, thus suppressing this additional neutron-production channel.

Based on simulation optimization results, we selected high-purity quartz (SiO$_2$) as the container material due to its negligible neutron scattering/absorption and good mechanical strength. The cylindrical vessel has an outer diameter of 80~mm and a total height of 150~mm, and is filled with D$_2$O. To experimentally verify the optimal source position predicted by the simulation (i.e., 1~cm from the forward end face for maximum forward neutron yield), we designed an axial adjustment mechanism using a manual linear feedthrough (CF16 flange, 100~mm stroke) driven by a handwheel.
\begin{figure}[htbp]
\centering
\includegraphics[width=1.0\linewidth]{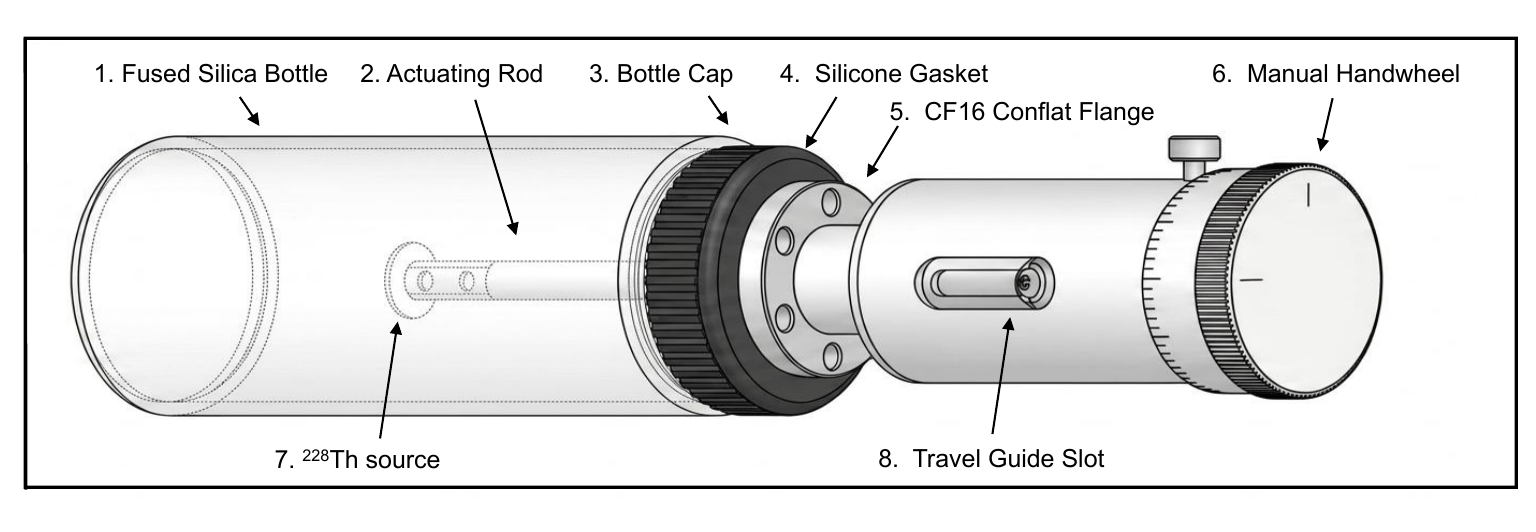} 
\caption{Schematic diagram of the optimized $^{228}$Th--D$_2$O photoneutron source assembly. Key components: 1, fused silica bottle; 2, actuating rod; 3, bottle cap; 4, silicone gaskets; 5, CF16 Conflat flange; 6, manual handwheel; 7, $^{228}$Th source capsule; 8, travel guide slot. The source-to-bottom distance (between the capsule and the inner bottom face) is adjustable, allowing experimental verification of the optimal axial position.}
\label{fig:source_schematic}
\end{figure}

The encapsulation procedure was as follows: the electrodeposited $^{228}$Th disk was first placed in a cylindrical plastic container and finally the whole assembly was fixed to the end of the feedthrough rod with epoxy resin. By rotating the handwheel, the source can be moved axially inside the vessel, enabling precise positioning while ensuring physical isolation from the D$_2$O. A schematic diagram of the assembled source is shown in Fig.~\ref{fig:source_schematic}.

\section{Neutron yield measurement}
\label{sec:yield_method}

Forward neutron emission was characterized using a single $^{3}$He proportional counter, which takes advantage of the large thermal-neutron capture cross section of $^{3}$He through the reaction: 

\begin{equation}
{}^{3}\mathrm{He} + n
\rightarrow {}^{3}\mathrm{H} + p + 764~\mathrm{keV}.
\label{eq:he3_capture}
\end{equation}

The kinetic energy released in Eq.~\eqref{eq:he3_capture} is shared between the proton and the triton, producing a well-defined full-energy peak that is clearly separated from the $\gamma$-ray background in the pulse-height spectrum.

The proportional counter has a diameter of 5.0\,cm, an effective sensitive length of 170\,mm, and is filled with 1\,bar of $^{3}$He gas. A cylindrical polyethylene (PE) moderator with a thickness of 4\,cm surrounds the detector. The moderator thickness was optimized using Geant4 simulations to maximize the detection efficiency for neutrons in the 0.17--0.22\,MeV energy range. Fast neutrons are moderated to thermal energies within the PE before being captured in the $^{3}$He gas.

\begin{figure}[htbp]
\centering
\includegraphics[width=0.7\linewidth]{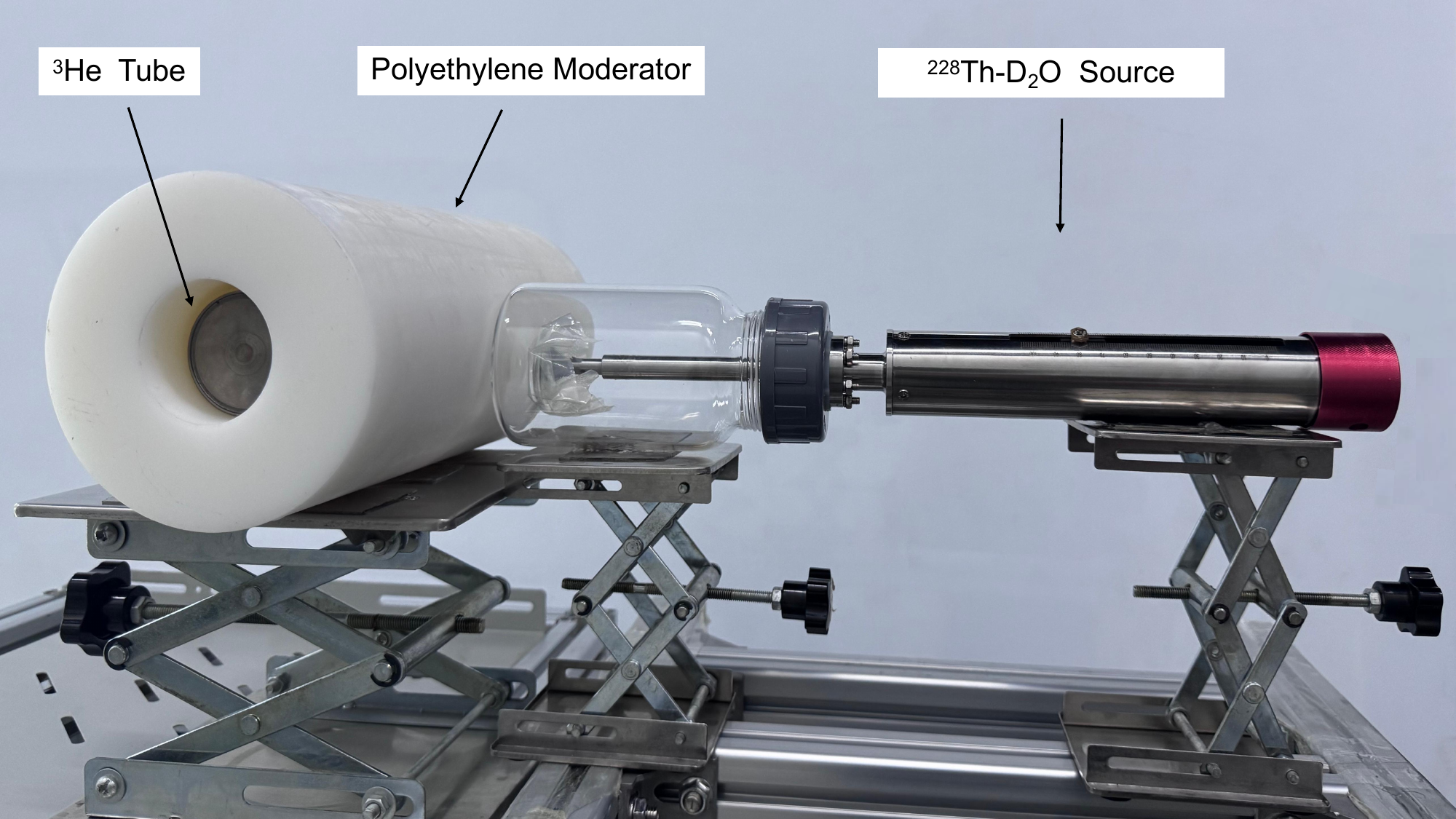}
\caption{Experimental setup for the forward neutron yield measurement. From left to right are the $^{3}$He proportional counter, the PE moderator, and the $^{228}$Th--D$_2$O source. The $^{3}$He tube has a diameter of 5~cm, a total length of 200~mm, and an effective sensitive length of 170~mm. The PE moderator has a density of 0.95~g/cm$^3$ and a thickness of 4~cm. The relative position between the source and detector is adjusted using a lift stage mounted on an aluminum support platform. The entire setup is elevated approximately 1.5~m above the concrete floor to reduce room-return neutrons. During the measurement, the bottom face of the D$_2$O vessel is placed in contact with the PE moderator.}
\label{fig:Bottom-exp}
\end{figure}

The detector was biased at +1200\,V using a high-voltage supply. The signals were processed by a charge-sensitive preamplifier, a spectroscopy amplifier, and a multichannel analyzer (MCA). The setup is shown in Fig.~\ref{fig:Bottom-exp}. For each source position from 0 to 5\,cm, data were first collected with the $^{228}$Th--D$_2$O source and then repeated under otherwise identical conditions with H$_2$O replacing D$_2$O to determine the $\gamma$-induced background. Each measurement lasted approximately 20\,h.

The Geant4 simulation was used to determine the absolute detection efficiency of the measurement system using the exact experimental geometry. The detection efficiency was defined as the ratio of the number of neutrons captured in the sensitive volume of the $^{3}$He detector to the number of neutrons forward-emitted from the forward end face of the D$_2$O target toward the detector. For the optimal source position (1~cm), the simulated absolute detection efficiency is 7.63\%. This efficiency includes neutron transport, moderation, and capture processes in the detector system.

\begin{figure}[htbp]
\centering
\includegraphics[width=0.6\linewidth]{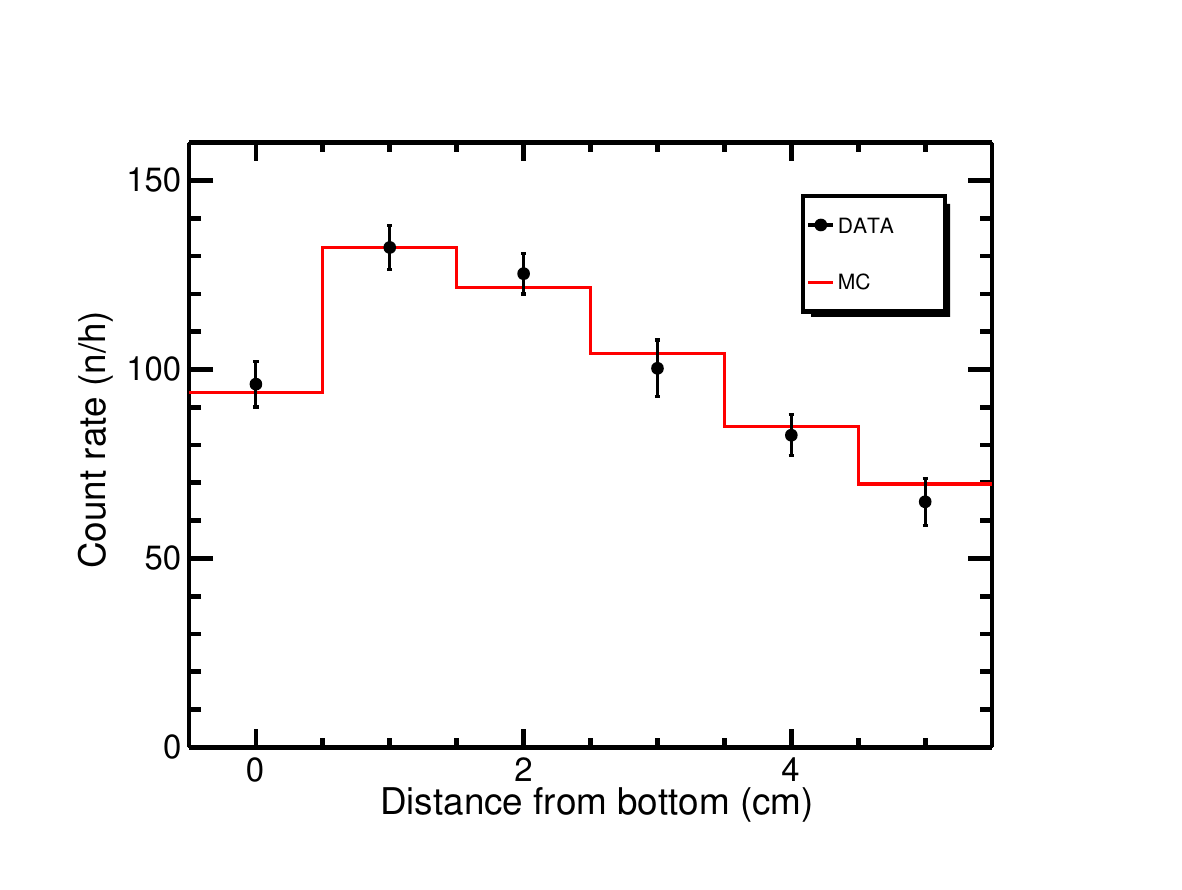}
\caption{Net count rate (n/h) as a function of distance from the bottom face.
The black circles with error bars represent the measured net count rates,
while the red stepped line denotes the Geant4 simulation results normalized
to the data at 1\,cm. Both datasets exhibit a consistent trend, with a peak
observed near 1\,cm.}
\label{fig:Bottom_Net_vs_MC}
\end{figure}

The same simulation was also used to calculate the relative variation of the neutron detection rate as a function of source position. Figure~\ref{fig:Bottom_Net_vs_MC} compares the measured net neutron count rates with the Geant4 prediction after normalizing both distributions at the 1~cm position. The agreement supports the detector model and indicates that the forward neutron emission is maximized when the radioactive source is positioned 1~cm from the forward end face of the D$_2$O target.

Since only the neutron emission rate at the optimal 1~cm position is used for the absolute neutron-emission-rate determination, the systematic uncertainty was evaluated exclusively for this configuration. Individual uncertainty contributions were estimated by varying the corresponding geometric or material parameters within their expected manufacturing tolerances or experimental uncertainties in the Geant4 simulation while keeping all other parameters unchanged. The resulting variation in the predicted $^{3}$He capture counts was taken as the systematic uncertainty associated with each parameter.

The systematic uncertainties were evaluated separately for the neutron source and the $^{3}$He detection system. The source-related uncertainties considered for the detection-efficiency calculation include the D$_2$O filling condition, D$_2$O density, source position, and quartz vessel wall thickness. In order of decreasing impact, the D$_2$O filling condition is the dominant source-related uncertainty. In the source-filling procedure, a small amount of residual air cannot be completely removed from the quartz vessel after filling with D$_2$O. The residual air is expected to accumulate at the uppermost part of the vessel after the source assembly is positioned, resulting in an unfilled gap above the D$_2$O surface. Based on the filling procedure and the dimensions of the vessel, a maximum residual air gap of 1~cm was considered as a conservative estimate of the filling uncertainty. The corresponding 3.27\% variation is defined as the absolute relative difference between the simulated $^{3}$He capture counts for the nominal fully filled geometry and for the geometry including the 1~cm residual air gap, i.e., $|N_{\rm gap}-N_{\rm nominal}|/N_{\rm nominal}\times100\%$. The next largest contributions are from the D$_2$O density and source position, with variations of 1.24\% and 0.51\%, respectively. The effects of the $^{228}$Th source activity and quartz vessel wall thickness are negligible within the investigated range.

The detector-related uncertainties include the $^{3}$He gas pressure and purity, the effective sensitive length of the proportional counter, the titanium tube-wall thickness, and the PE moderator properties (density and thickness). In order of decreasing impact, the corresponding variations in the simulated capture counts are 0.99\% (PE thickness), 0.52\% (Ti tube-wall thickness), 0.51\% ($^{3}$He gas pressure), 0.39\% (effective sensitive length), 0.06\% (PE density), and $<0.01\%$ ($^{3}$He gas purity). The effect of the $^{3}$He gas purity is negligible within the investigated range.

The individual contributions are summarized in Table~\ref{tab:sys}. The $^{228}$Th source activity is not included in the systematic uncertainty of the simulated detection efficiency, because the efficiency is determined independently of the absolute source activity. The individual efficiency-related contributions are added in quadrature, giving a total systematic uncertainty of 3.76\%.

\begin{table}[htbp]
\centering
\caption{Systematic uncertainties in the simulated $^{3}$He detection efficiency for the optimal source position (1~cm).}
\label{tab:sys}
\begin{tabular}{lc}
\toprule
Source of uncertainty & Relative uncertainty (\%) \\
\midrule
\multicolumn{2}{l}{\textit{Source-related}} \\
D$_2$O filling condition (1~cm air gap) & 3.27 \\
D$_2$O density & 1.24 \\
Source position & 0.51 \\
Quartz vessel wall thickness & $<0.01$ \\
\midrule
\multicolumn{2}{l}{\textit{Detector-related}} \\
PE thickness & 0.99 \\
Ti tube wall thickness & 0.52 \\
$^{3}$He gas pressure & 0.51 \\
Effective sensitive length & 0.39 \\
PE density & 0.06 \\
$^{3}$He gas purity & $<0.01$ \\
\midrule
Total & 3.76 \\
\bottomrule
\end{tabular}
\end{table}

At the optimal source position, the measured count rate with the $^{228}$Th--D$_2$O source is $R_{\mathrm{D_2O}} = 285.0 \pm 4.2~\mathrm{h}^{-1}$, while the corresponding background count rate measured with the $^{228}$Th--H$_2$O source is $R_{\mathrm{H_2O}} = 152.7 \pm 4.0~\mathrm{h}^{-1}$. The background-subtracted count rate is therefore
\begin{equation}
R_{\mathrm{net}}
= R_{\mathrm{D_2O}} - R_{\mathrm{H_2O}}
= 132.3 \pm 5.8~\mathrm{h}^{-1}.
\label{eq:net_rate}
\end{equation}
where the uncertainty is statistical only and is obtained by adding the statistical uncertainties of the two independent measurements in quadrature. The net rate represents the count rate attributable to neutrons produced by the $^{228}$Th--D$_2$O source after subtraction of the background measured with H$_2$O.

Combining the net count rate with the Geant4-derived absolute detection efficiency of 7.63\%, the forward neutron emission rate is determined to be
\begin{equation}
R_{\mathrm{n,forward}}
= 1734 \pm 76~(\mathrm{stat.}) \pm 65~(\mathrm{sys.})~\mathrm{h}^{-1}.
\label{eq:forward_rate}
\end{equation}

\section{Neutron energy measurement}
\label{sec:energy_measurement}

The accurate determination of neutron energy spectra is fundamental to neutron physics research. In practical measurement scenarios, neutron energies often span a broad range, from thermal to fast neutrons, which imposes stringent requirements on the energy response of the detection system. The BSS is one of the most widely used and well-established tools for neutron spectrometry in both radiation protection monitoring and fundamental physics experiments. Its popularity stems from its broad energy coverage (typically from thermal to MeV), approximately isotropic response, and operational simplicity \cite{Thomas2002,Bramblett1960}.

The operating principle of the BSS does not rely on the direct measurement of individual neutron energies; instead, it is based on a ``moderation--detection'' process. A thermal-neutron-sensitive detector is placed at the center of a spherical moderator made of a hydrogenous material, typically high-density PE; measurements are repeated with moderator spheres of different diameters. Incident neutrons of different energies undergo varying degrees of moderation when traversing these spheres. Small-diameter spheres have their maximum response at low neutron energies, whereas large-diameter spheres shift the response toward higher energies. By successively replacing the moderator spheres, one obtains a set of count rates, each representing a different energy-dependent response of the spectrometer to the incident neutron spectrum. This count-rate vector does not itself constitute the spectrum; rather, it represents the incident neutron spectrum weighted by the response functions of the individual moderator configurations. The actual neutron spectrum is then recovered using mathematical unfolding algorithms---such as the maximum-entropy method (MAXED)---with a response matrix calculated in advance by Monte Carlo simulations \cite{Reginatto2002,Matzke2002}.

The BSS system used in this study was provided by the National Institute of Metrology, China. The central detector is a spherical $^{3}\mathrm{He}$ proportional counter filled with $^{3}\mathrm{He}$ gas at a pressure of 2.5~bar. This detector exploits the exceptionally high thermal-neutron absorption cross section of the $^{3}\mathrm{He}(n,p)^{3}\mathrm{H}$ reaction, approximately $5330~\mathrm{barn}$ at $2.53 \times 10^{-2}~\mathrm{eV}$, and records neutrons by measuring the ionization pulses generated by the reaction products. In addition, the proportional counter exhibits negligible sensitivity to $\gamma$ rays, thereby effectively suppressing interference from the ambient $\gamma$-ray background. The thermal-neutron response of the detector was calibrated at the Physikalisch-Technische Bundesanstalt (PTB), Germany, yielding a response of $(3.21 \pm 0.3)~\mathrm{cm}^{2}$ \cite{LuszikBhadra2014PTB,Wang2022NuclearPower}. From this value, the $^{3}\mathrm{He}$ atomic number density was determined to be $6.07 \times 10^{19}~\mathrm{cm}^{-3}$. The relative uncertainty in the response function arising from the $^{3}\mathrm{He}$ pressure was estimated to be approximately 1\%.

To obtain complete spectral information for the $^{228}$Th--D$_2$O source, the system was equipped with a full set of PE moderator spheres. A total of nine configurations were used: the bare detector (without any moderator, sensitive primarily to thermal and epithermal neutrons) and PE spheres with outer diameters of 3, 3.5, 4, 4.5, 5, 6, 7, and 8 inches. 

\begin{figure}[htbp] 
  \centering   
  \begin{subfigure}[b]{0.50\textwidth}    
    \centering    
    \includegraphics[width=\linewidth]{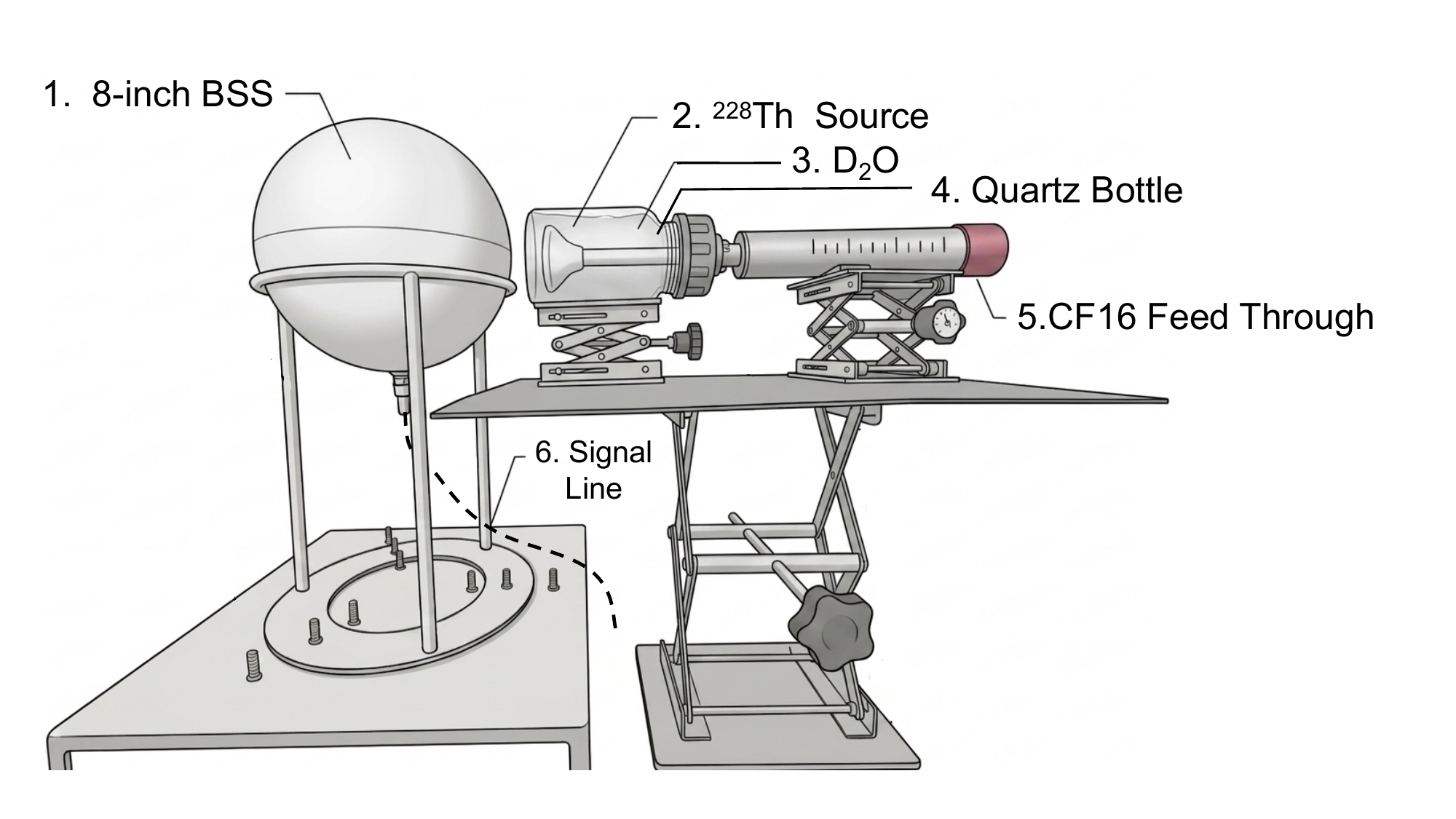}    
    \caption{}      
    \label{fig:ThD-bonner-exp}  
  \end{subfigure}  
  \hfill   
  \begin{subfigure}[b]{0.48\textwidth}    
    \centering       
    \includegraphics[width=\linewidth]{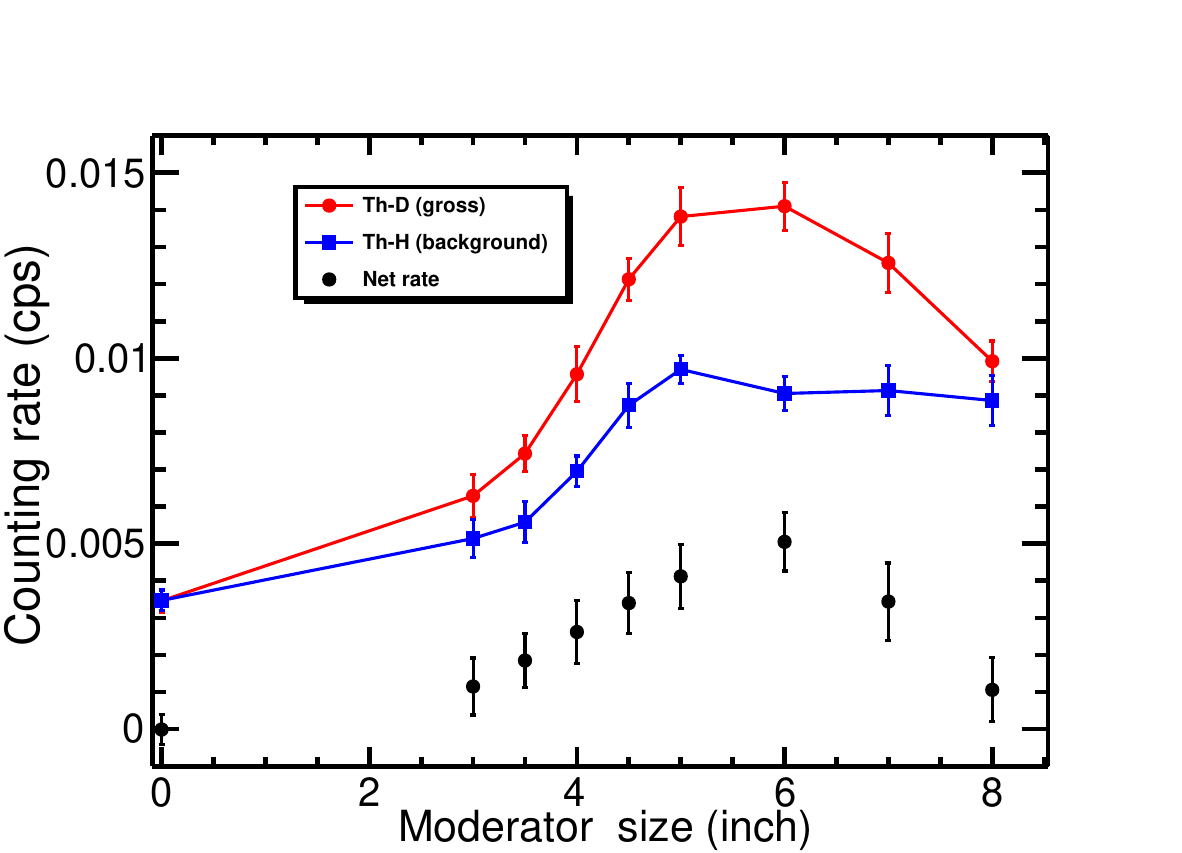}    
    \caption{}      
    \label{fig:ThD-all-rates}   
  \end{subfigure}    
\caption{Experimental setup and measurement results. (a) Schematic of the experimental arrangement, generated with Google Gemini and checked by the authors against the actual apparatus. Numbered components: 1, 8-inch Bonner sphere; 2, $^{228}$Th source; 3, D$_2$O; 4, quartz bottle; 5, CF16 feedthrough; 6, signal line. The spheres are mounted on an aluminum frame facing the source, and the source capsule is held 1~cm from the forward end face of the SiO$_2$ flask. (b) Count rate versus moderator diameter. The red points correspond to the $^{228}$Th--D$_2$O source, the blue points to the background measurement with H$_2$O substituted for D$_2$O, and the black points to the net count rate obtained after subtracting the H$_2$O background from the D$_2$O data.}
  \label{fig:combined-setup}    
\end{figure}

This range covers the moderation requirements from low-energy thermal neutrons to high-energy fast neutrons, ensuring accurate spectrum determination.
The experimental arrangement is shown in Fig.~\ref{fig:combined-setup}(a). To minimize neutron scattering from the surroundings, the entire assembly was elevated 1.5~m above the concrete floor, and the distance from the source center to the detector center was fixed at 13~cm. For both the signal and background measurements, all nine moderator configurations were measured under the same geometry. The signal processing chain comprised a preamplifier, a spectroscopy amplifier, and a multichannel analyzer, and the detector high voltage was supplied by a NIM-standard high-voltage module.

For the signal measurement, the nine configurations were measured sequentially from the bare detector to the 8-inch sphere with the $^{228}$Th--D$_2$O source. The D$_2$O was then replaced with H$_2$O, and the same sequence was repeated for the background measurement with the $^{228}$Th--H$_2$O source. The net neutron count rate for each configuration was obtained by subtracting the background count rate from the signal count rate. The measured and net count rates are shown in Fig.~\ref{fig:combined-setup}(b). The $^{228}$Th--D$_2$O source served as the signal source, while the $^{228}$Th--H$_2$O source served as the background reference. In D$_2$O, deuterium nuclei undergo $(\gamma,n)$ reactions induced by $\gamma$ rays from the $^{228}$Th decay chain, whereas in H$_2$O, hydrogen does not produce neutrons at these energies. The H$_2$O source therefore provides a well-matched background for subtracting contributions from direct $\gamma$ rays, source encapsulation, ambient neutrons, and room-scattered neutrons.

The net count rate reaches a maximum at the 6-inch moderator diameter. This is consistent with the known peak sensitivity of the 6-inch PE sphere at neutron energies around 0.2~MeV and with the expected neutron spectrum of the $^{228}$Th--D$_2$O source, which is known to peak in this energy region. This consistency supports the reliability of the experimental setup and the subsequent unfolding analysis.
The unfolding of the neutron spectrum was performed using the MXD\_FC31 code, which is part of the UMG~3.1 (Unfolding with MAXED and GRAVEL) software package developed by the PTB~\cite{UMG31}. This algorithm is based on the maximum-entropy principle and has been specifically optimized for BSS data~\cite{Reginatto2002}. In contrast to the default response-function library distributed with the UMG software, the response functions used in this study were re-modeled and simulated using the Geant4 Monte Carlo toolkit~\cite{Agostinelli2003}. A broad parallel neutron beam was used in the simulations, with the incident direction aligned along the symmetry axis of the detector. Simulations were performed for all configurations, from the bare detector through to the 8-inch sphere. The Geant4 spectrum was used as the \textit{a priori} (default) spectrum required by the maximum-entropy algorithm, following the practice of using calculated spectra as prior information in BSS unfolding~\cite{Thomas2020}. In this formulation the prior enters only as the reference spectrum of the entropy functional, and the solution is still required to reproduce the nine measured count rates.

\begin{figure}[htbp]
\centering
\includegraphics[width=0.60\linewidth]{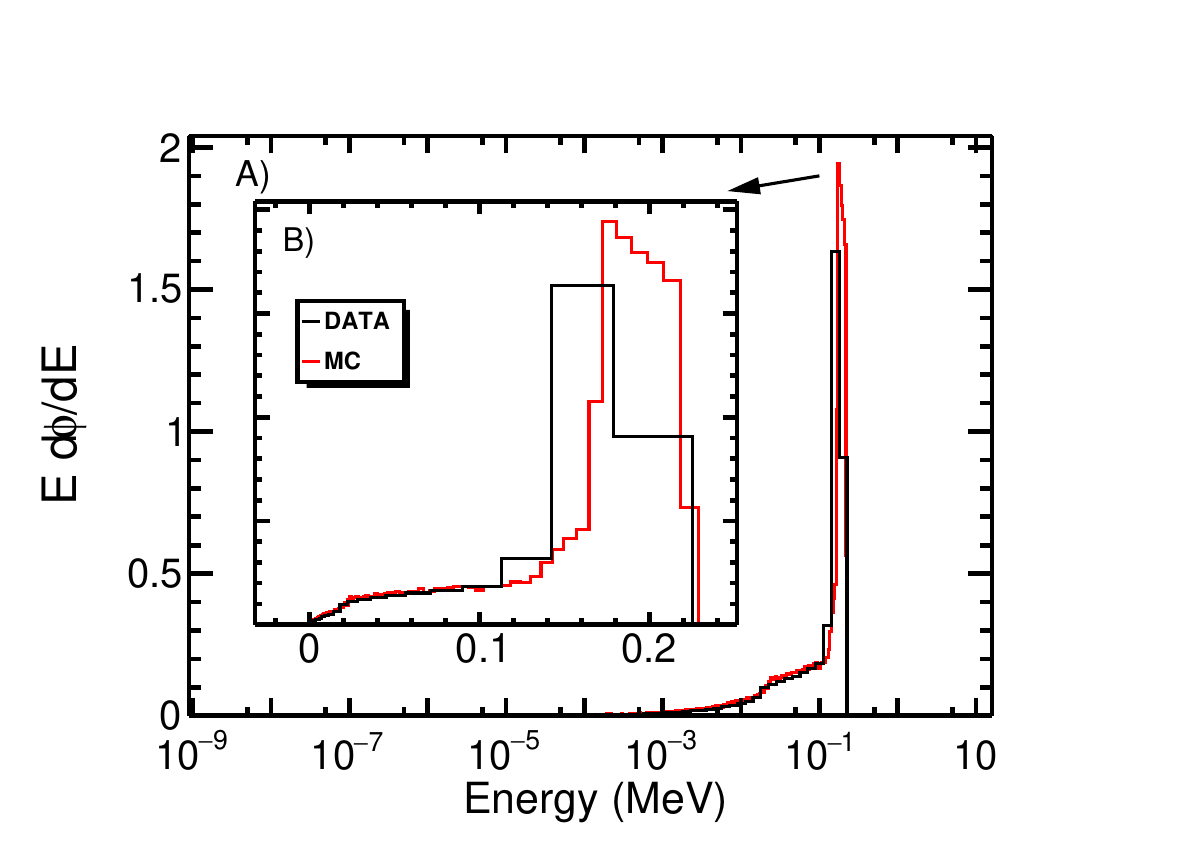}
\caption{Energy spectra of the $^{228}$Th--D$_2$O source obtained from experimental unfolding (black, DATA) and Geant4 simulation (red, MC). Both spectra are normalized such that the area under $d\phi/dE$ is unity, i.e., $\int (d\phi/dE)\,dE = 1$; the vertical axis shows $E\,d\phi/dE$. The main panel (A) uses a logarithmic horizontal scale, while the inset (B) shows the 0--0.24~MeV region on a linear scale. The mean energies are 0.125~MeV for DATA and 0.128~MeV for MC.}
\label{fig:ThD-energy}
\end{figure}

The unfolded spectra are presented in Fig.~\ref{fig:ThD-energy}. Panel (A) uses a logarithmic horizontal scale, while panel (B) uses a linear scale. The experimental and simulated spectra show similar overall shapes, with both exhibiting a prominent feature around 0.2~MeV. A quantitative assessment of the statistical significance of any discrepancy is not attempted here because the figure does not show an uncertainty band for the unfolded spectrum. The overall spectral shape is consistent with expectations, as the excess energy of the 2.615~MeV $\gamma$ ray above the deuteron photodisintegration threshold is limited.

From the reaction kinematics, the 2.615~MeV $\gamma$ ray exceeds the deuteron binding energy (2.224~MeV) by only approximately 0.391~MeV. Consequently, the emitted neutron energy is governed by the two-body kinematics of the $^{2}$H($\gamma$,n)p reaction and remains entirely in the sub-MeV region. The full unfolded spectra have mean energies of 0.125~MeV for DATA and 0.128~MeV for MC, as shown in Fig.~\ref{fig:ThD-energy}. Within these spectra, the prominent forward-emission feature is concentrated in the 0.17--0.22~MeV region, hereafter referred to as the peak region.

The gradually decreasing intensity observed across the peak region is primarily governed by the geometry-dependent escape probability rather than the intrinsic reaction kinematics. Detailed Geant4 track-by-track analysis shows that neutrons within this energy region exhibit a clear correlation between their kinetic energy and momentum direction. Neutrons with energies closer to the upper end of the peak region predominantly possess momentum components opposite to the forward direction and therefore must traverse a longer distance through the D$_2$O moderator before escaping from the forward surface. Consequently, they experience a higher probability of elastic scattering and attenuation. In contrast, lower-energy neutrons within the peak region are preferentially emitted toward the forward direction and escape more readily.

The interplay between the initial photoneutron distribution and the energy-dependent escape probability naturally gives rise to the observed gradually decreasing spectral structure in the peak region. A detailed Geant4 track-by-track analysis further reveals the physical origin of the remaining spectral structures. Besides the primary photoneutron distribution, elastic scattering on deuterium and oxygen nuclei redistributes the neutron energy and modifies the escaping spectrum. Owing to the much larger kinematically allowed recoil energy of deuterium, neutron scattering on D produces a broad continuum extending approximately from 0.02 to 0.22~MeV. In contrast, scattering on oxygen mainly contributes to the higher-energy shoulder between approximately 0.17 and 0.22~MeV because of its much smaller recoil-energy transfer. Neutrons below approximately 0.02~MeV are predominantly produced after multiple scattering events inside the moderator before escaping from the source. The measured neutron spectrum can therefore be understood as the superposition of the primary photoneutron spectrum and the energy redistribution introduced by neutron scattering on deuterium, neutron scattering on oxygen, and multiple scattering inside the D$_2$O moderator. The maximum recoil energies extracted from the Geant4 track analysis are consistent with the theoretical prediction of two-body elastic-scattering kinematics, $E_R^{\rm max}=4AE_n/(A+1)^2$, yielding approximately 0.196~MeV for deuterium ($A=2$) and 0.049~MeV for oxygen ($A=16$) for a 0.22~MeV incident neutron, providing strong support for this physical interpretation.


\section{Simulation of the PandaX-20T calibration}
\label{sec:pandax_sim}

Based on the experimentally characterized properties of the $^{228}$Th--D$_2$O photoneutron source described above, a dedicated Monte Carlo study was performed to evaluate its calibration performance in the PandaX-20T LXe TPC. PandaX-20T is the next-generation detector of the PandaX dark matter experiment and is expected to operate at the China Jinping Underground Laboratory (CJPL) in 2027~\cite{PandaXxT2025}. The detector consists of a cylindrical dual-phase LXe TPC with an approximate diameter of 2.0~m and a height of 2.5~m, corresponding to a LXe target mass of about 20 tons. The simulation was performed using the PandaX Geant4-based Monte Carlo framework BambooMC~\cite{Chen2021BambooMC}.

The deployment position of the neutron source was first optimized. Identical neutron sources with the same energy spectrum and activity were placed at the top and lateral positions of the LXe TPC, respectively, and the number of NR events deposited in the active LXe volume was compared. The simulation results show that the top-position configuration produces a higher NR event yield than the side-position configuration, mainly due to the more favorable source-to-target geometry. Therefore, the neutron source was deployed at the top of the PandaX-20T detector for subsequent optimization and performance evaluation.

\begin{figure}[htbp]
    \centering
    \makebox[\textwidth][c]{\hspace*{1cm}\includegraphics[width=0.75\textwidth]{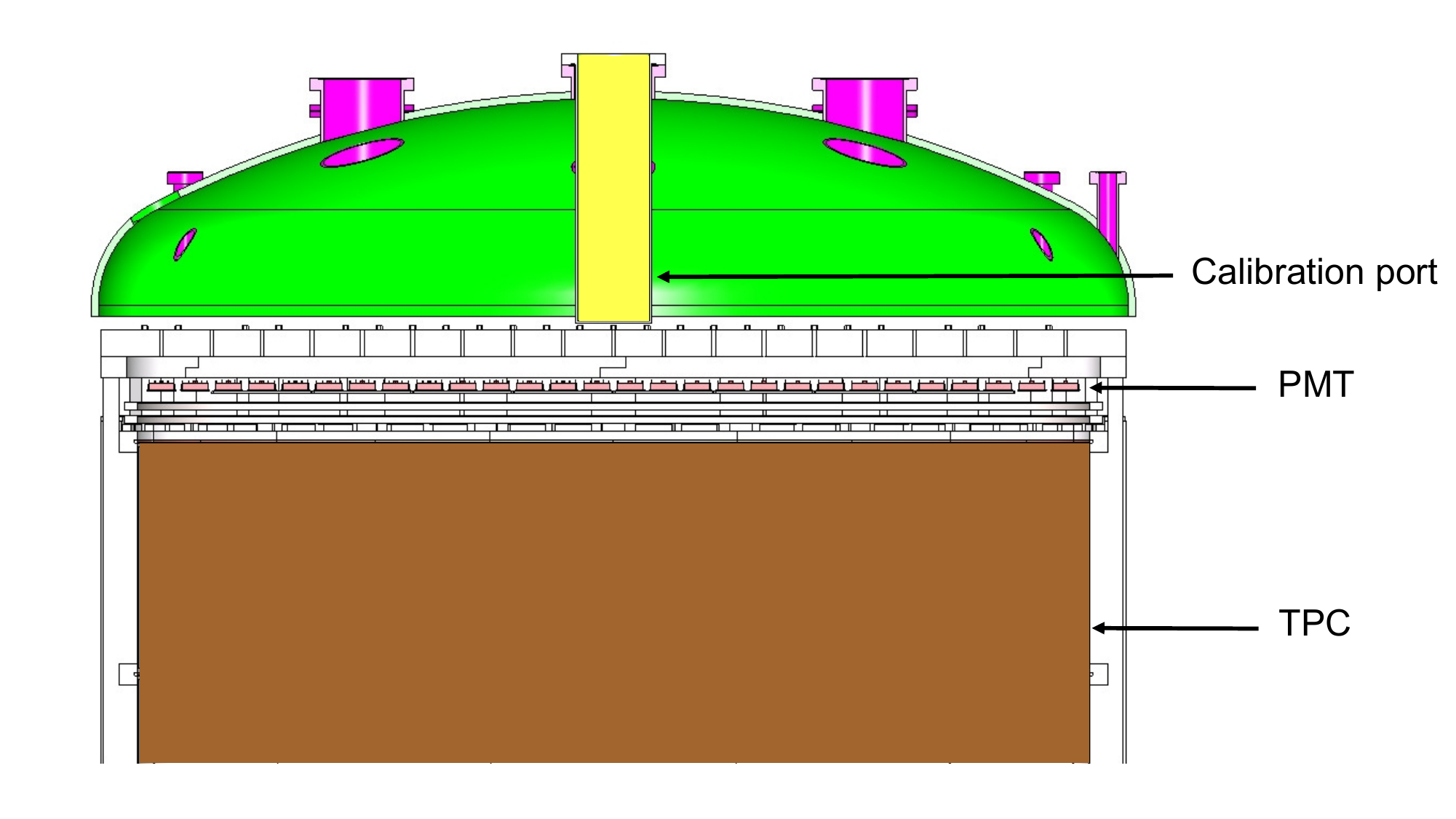}}
    \caption{Schematic view of the upper section of the PandaX-20T inner vessel, illustrating the relative positions of the top calibration port, the photomultiplier tube (PMT) array, and the time projection chamber (TPC).}
    \label{fig:pandax_geometry}
\end{figure}

A cylindrical calibration port with a diameter of approximately 150~mm is reserved at the top of PandaX-20T for external radioactive sources, as illustrated in Fig.~\ref{fig:pandax_geometry}. For the $^{228}$Th--D$_2$O photoneutron source, the shielding geometry was optimized to maximize the ratio of NR to ER events ($N_{NR}/N_{ER}$), while the resulting event rate was subsequently checked against the data acquisition (DAQ) capability of the detector. Since the source emits 2.615~MeV $\gamma$ rays isotropically, both lateral and forward $\gamma$-ray leakage were considered. Tungsten shielding structures were therefore introduced around the source for the NR calibration configuration, aiming to suppress the $\gamma$-induced ER background while preserving the neutron calibration efficiency.

The lateral shielding was optimized using a tungsten annular shield surrounding the D$_2$O volume. Its outer diameter was fixed at 150~mm to match the diameter of the calibration port, while its inner diameter was varied from 150~mm (no tungsten shielding) to 20~mm. The volume inside the inner diameter was filled with D$_2$O. The forward shielding was optimized using a cylindrical tungsten shield placed between the neutron source and the LXe target. Its diameter was fixed at 150~mm, matching the diameter of the calibration port, while its axial thickness was varied to investigate its influence on the $N_{NR}/N_{ER}$ ratio.

\begin{figure}[htbp]
    \centering
    \includegraphics[width=0.9\textwidth]{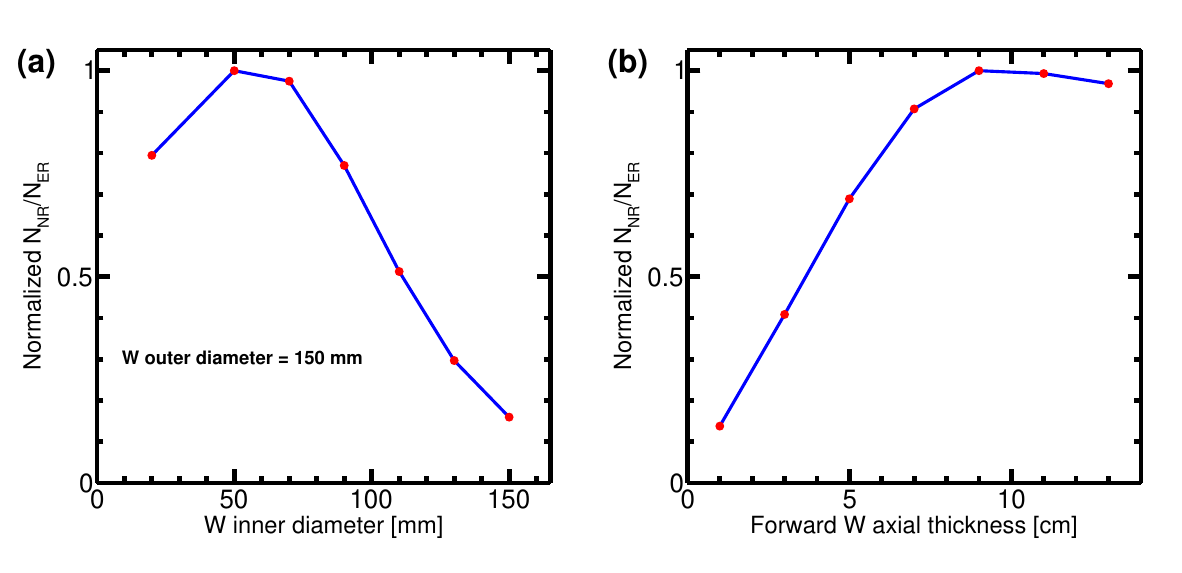}
    \caption{Optimization of the tungsten (W) shielding geometry based on the simulated NR-to-ER ratio. (a) $N_{NR}/N_{ER}$ as a function of the inner diameter of the tungsten annulus with a fixed outer diameter of 150~mm. (b) $N_{NR}/N_{ER}$ as a function of the forward tungsten axial thickness with a fixed lateral shielding configuration.}
    \label{fig:shielding_opt}
\end{figure}

The optimization results are summarized in Figs.~\ref{fig:shielding_opt}(a) and~\ref{fig:shielding_opt}(b). The maximum $N_{NR}/N_{ER}$ ratio is achieved with a lateral tungsten annulus having an inner diameter of 50~mm and an outer diameter of 150~mm. With this lateral shielding configuration, the forward tungsten thickness was further optimized to balance the suppression of $\gamma$-induced ER events and the transmission of calibration neutrons. The $N_{NR}/N_{ER}$ ratio reaches its maximum at a forward tungsten axial thickness of approximately 9~cm, indicating that this configuration provides the most favorable balance between background suppression and neutron transmission among the investigated configurations. Therefore, the final shielding configuration consists of a lateral tungsten annulus with an inner diameter of 50~mm and an outer diameter of 150~mm, together with a forward cylindrical tungsten shield with a diameter of 150~mm and an axial thickness of 9~cm. It should be noted that these dimensions represent the optimized shielding geometry in the Monte Carlo study. The final implementation may require minor adjustments according to engineering constraints and the as-built experimental setup, while the overall shielding optimization strategy remains unchanged.

Using the optimized shielding configuration, the calibration performance of the $^{228}$Th--D$_2$O neutron source in PandaX-20T was further evaluated. The unfolded neutron spectrum obtained from the BSS measurement was used to specify the energy dependence of the incident neutron field, while its absolute normalization was set independently by the experimentally measured forward neutron emission rate. In this treatment, the BSS result provides the spectral shape and the $^{3}$He-counter measurement provides the absolute forward-emission normalization. This separation should be understood as an explicit shape--normalization treatment rather than as assuming that the BSS and forward-emission measurements have identical absolute definitions. The $\gamma$-ray emission rate was normalized according to the measured $^{228}$Th activity and the $^{208}$Tl branching ratio.

\begin{figure}[htbp]
    \centering
    \includegraphics[width=0.9\textwidth]{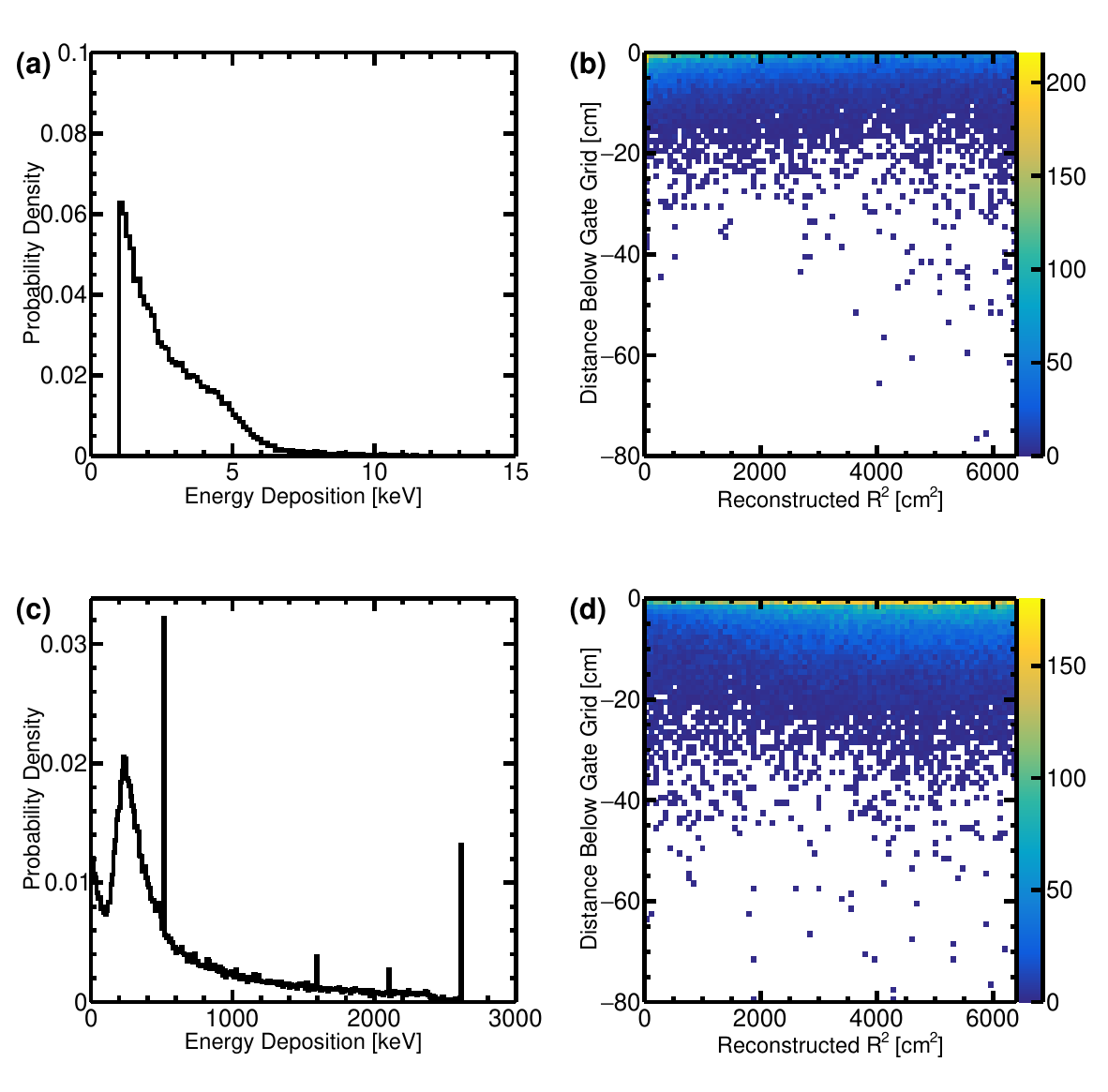}
    \caption{Simulated energy deposition and spatial distributions induced by neutrons and the 2.615~MeV $\gamma$ ray from the $^{228}$Th--D$_2$O source in the PandaX-20T detector. (a) Energy deposition spectrum induced by neutrons. The sharp threshold at 1~keV corresponds to the $E_{\rm NR}>1~\mathrm{keV}_{\mathrm{nr}}$ selection, and the cutoff at approximately 6~keV is consistent with the maximum elastic recoil energy of 0.22~MeV neutrons on $^{131}$Xe, $E_R^{\max}=4AE_n/(A+1)^2\approx6~\mathrm{keV}$. (b) Corresponding spatial distribution in the reconstructed $R^2$--$Z$ plane for neutron-induced events. (c) Energy deposition spectrum induced by the 2.615~MeV $\gamma$ ray alone. The main peaks correspond to the 0.511~MeV full-energy peak, the double-escape peak of the 2.615~MeV line ($\sim$1.59~MeV), the single-escape peak ($\sim$2.10~MeV), and the full-energy peak ($\sim$2.615~MeV). The continuum mainly originates from Compton scattering of the 2.615~MeV $\gamma$ ray. (d) Corresponding spatial distribution in the reconstructed $R^2$--$Z$ plane for 2.615~MeV $\gamma$-induced events.}
    \label{fig:deposition_dist}
\end{figure}

The simulated NR energy spectrum and spatial distribution in the PandaX-20T LXe target are shown in Fig.~\ref{fig:deposition_dist}(a) and (b), while the corresponding ER energy spectrum and spatial distribution are presented in Fig.~\ref{fig:deposition_dist}(c) and (d). The simulation applies the NR energy threshold and single-scatter event selection, retaining events with $E_{\rm NR}>1~\mathrm{keV}_{\mathrm{nr}}$ and a single energy-deposition site. As shown in Fig.~\ref{fig:deposition_dist}(a), the NR spectrum exhibits a sharp threshold at 1~keV due to the selection condition and decreases with increasing energy, with a cutoff at approximately 6~keV. This cutoff is consistent with the kinematic limit for 0.22~MeV neutrons elastically scattering on Xe, $E_R^{\max}=4AE_n/(A+1)^2\approx6~\mathrm{keV}$, confirming that the source produces NRs concentrated in the few-keV region. The neutron-induced NR spectrum demonstrates that the $^{228}$Th--D$_2$O source provides a low-energy NR population in the energy region relevant for dark matter searches, where accurate characterization of the LXe response is essential~\cite{PhysRevC.81.025808,PhysRevC.84.045805,Aprile2018NR,PhysRevLett.123.231106}. Figure~\ref{fig:deposition_dist}(c) shows the energy deposition spectrum for the 2.615~MeV $\gamma$ ray alone. The main peaks correspond to the 0.511~MeV full-energy peak, the double-escape peak of the 2.615~MeV line ($\sim$1.59~MeV), the single-escape peak ($\sim$2.10~MeV), and the full-energy peak ($\sim$2.615~MeV). The continuum mainly originates from Compton scattering of the 2.615~MeV $\gamma$ ray. The spatial distributions of both NR and ER events illustrate the neutron and $\gamma$-ray transport, scattering, and energy deposition characteristics inside the LXe target, providing information on the calibration coverage and event distribution throughout the detector volume.

For the optimized shielding configuration, the simulated $\gamma$-induced event rate in the detector is approximately 30~Hz, which is compatible with the expected DAQ bandwidth. After applying the NR selection criteria of $E_{\rm NR}>1~\mathrm{keV}_{\mathrm{nr}}$ and requiring single-scatter events, the expected neutron-induced NR event rate is approximately 4.60~events~h$^{-1}$. An exposure of approximately 220~h would therefore provide about 1000 neutron-induced NR events in the PandaX-20T detector. Such a calibration sample provides sufficient statistics for studying the low-energy NR response of the LXe target.

It should be noted that the optimized shielding configuration is specifically designed for low-energy NR calibration by suppressing the accompanying $\gamma$-ray flux. The ER calibration capability discussed above therefore represents a separate operating mode in which the shielding configuration would need to be reduced or otherwise adjusted; the present simulation does not quantify the ER calibration performance of such an alternative configuration. In addition, the photoneutrons and the accompanying 2.615~MeV $\gamma$ rays originate from the same radioactive decay chain. The present simulation reports neutron-induced NR and $\gamma$-induced ER rates separately, but does not evaluate the probability that an accompanying $\gamma$ ray deposits energy in the same detector event window as a selected NR. This coincidence-induced ER contamination should be quantified in a future event-level simulation before claiming a specific NR sample purity.

These results demonstrate that the $^{228}$Th--D$_2$O photoneutron source can provide a controllable low-energy neutron calibration sample for PandaX-20T while remaining compatible with the detector DAQ bandwidth. With the optimized shielding configuration and the experimentally measured forward neutron emission rate, the source is expected to provide approximately $4.60~\mathrm{events~h^{-1}}$ neutron-induced NR events after the $E_{\rm NR}>1~\mathrm{keV}_{\mathrm{nr}}$ and single-scatter selections. The dedicated Monte Carlo study therefore demonstrates the feasibility of deploying this source for low-energy NR calibration in the PandaX-20T detector.

Although the simulated $\gamma$-induced ER rate is about $30~\mathrm{Hz}$, roughly four orders of magnitude higher than the neutron-induced NR rate, for the 2.615~MeV $\gamma$ line the energy deposition predominantly occurs in the MeV region, with almost no direct deposition in the few-keV NR region. Low-energy ER-like events can only arise from processes such as Compton scattering coincident with a low-energy X event, and such events are largely removed by the fiducial-volume and single-scatter selections. Similar considerations apply to material radioactivity: although abundant $\gamma$ rays are present, their contribution to the low-energy background inside the fiducial volume is very small. Therefore, the probability of high-energy ER leakage into the low-energy NR region is very small. A final validation based on the complete PandaX-20T detector simulation or data is still needed.

\section{Results and discussion}
\label{sec:results}

A long-lived, low-energy photoneutron source based on the $\mathrm{D}(\gamma,\mathrm{n})\mathrm{p}$ reaction was developed and systematically characterized. Geant4 simulations were performed to optimize the source geometry, showing that the forward neutron emission is maximized when the $^{228}$Th source is placed 1~cm from the forward end face of the D$_2$O volume. At this optimal position, the forward-emitted neutrons have kinetic energies below 1~MeV, with the measured and simulated spectra concentrated in the peak region. The simulated angular distribution of escaping neutrons is shown in Fig.~\ref{fig:rose_neutron}.

Experimental measurements using a moderated $^3$He proportional counter (Sect.~\ref{sec:yield_method}) provide a check of the simulation results. The measured net neutron count rates at different source positions agree well with the simulated spatial dependence (Fig.~\ref{fig:Bottom_Net_vs_MC}). At the 1~cm position, the measured net neutron count rate is given in Eq.~\eqref{eq:net_rate}. Combining this rate with the Geant4-calculated absolute detection efficiency of 7.63\%, we obtain the forward neutron emission rate given in Eq.~\eqref{eq:forward_rate}. The neutron energy spectrum measured with the BSS (Sect.~\ref{sec:energy_measurement}) and unfolded using the MAXED method with Geant4 response functions agrees well with the simulation (Fig.~\ref{fig:ThD-energy}). The emitted neutrons are concentrated in the peak region, consistent with the two-body kinematics of the $^2$H($\gamma$,n)p reaction~\cite{Marin1953}. The observed spectral shape results from the combined effects of reaction kinematics, elastic scattering on deuterium and oxygen, and the energy-dependent escape probability, as confirmed by track-by-track Geant4 analysis.

Compared with conventional $^{88}$Y--Be sources~\cite{Collar2013,Robinson2016}, this source provides a cleaner low-energy neutron spectrum without the additional $\sim$0.95~MeV neutron component and offers a substantially longer half-life (1.91~years versus 106.6~days). As summarized in Table~\ref{tab:comparison}, the neutron spectrum of the developed source is intrinsically matched to the NR energy region of interest for dark matter searches, since these neutrons can generate few-keV NRs through elastic scattering in LXe targets.

\begin{table}[htbp]
\centering
\small
\caption{Comparison of commonly used neutron calibration sources for LXe detector calibration. The listed properties are taken from Refs.~\cite{Mannhart1987,Marsh1995AmBe,Liskien1973DD,Collar2013,Robinson2016} and the present work.}
\label{tab:comparison}
\renewcommand{\arraystretch}{0.85}
\begin{tabular}{lccc}
\toprule
Source & Neutron energy & Half-life & Typical application \\
\midrule
$^{252}$Cf & Broad (0--10 MeV) & 2.65 y & Commissioning \\
AmBe & Broad (0--11 MeV) & 432 y ($^{241}$Am) & General calibration \\
D--D Generator & 2.45 MeV & N/A & Monoenergetic calibration \\
$^{88}$Y--Be & $\sim$0.15 + $\sim$0.95 MeV & 106.6 d & Low-energy NR calibration \\
$^{228}$Th--D$_2$O & 0.17--0.22 MeV & 1.91 y & Long-term low-energy NR calibration \\
\bottomrule
\end{tabular}
\end{table}

A dedicated Monte Carlo study for the PandaX-20T detector with the optimized tungsten shielding configuration predicts a neutron-induced NR event rate of approximately $4.60~\mathrm{events~h^{-1}}$ after applying the $E_{\rm NR}>1~\mathrm{keV}_{\mathrm{nr}}$ and single-scatter selections. The corresponding $\gamma$-induced ER rate is approximately $30~\mathrm{Hz}$, which satisfies the detector DAQ bandwidth requirement. An exposure of approximately $220~\mathrm{h}$ would accumulate about 1000 NR events. Although the ER rate is about four orders of magnitude higher than the NR rate, the probability of high-energy ER leakage into the low-energy NR region is very small.

In conclusion, the $^{228}$Th--D$_2$O photoneutron source developed in this work provides a stable and well-characterized neutron field for low-energy NR calibration. The measured neutron emission rate and energy spectrum are consistent with the Geant4 simulations, supporting the reliability of the optimized source design. Compared with conventional neutron calibration sources, this source provides a long-lived and intrinsically low-energy neutron spectrum suitable for few-keV NR studies in LXe detectors. The Monte Carlo study of PandaX-20T further indicates that the optimized source configuration can provide a sufficient number of calibration events for characterizing the low-energy NR response in future large-scale LXe experiments.

\section*{CRediT authorship contribution statement}

\textbf{Yueqiang Tian:} Conceptualization, Methodology, Software, Validation, Formal analysis, Investigation, Data curation, Writing -- original draft, Visualization.
\textbf{Chen Cheng:} Conceptualization, Methodology, Supervision, Writing -- review \& editing.
\textbf{Yuxin Cui:} Investigation, Writing -- review \& editing.
\textbf{Xunan Guo:} Investigation, Writing -- review \& editing.
\textbf{Zichao Guo:} Investigation, Writing -- review \& editing.
\textbf{Yao Tong:} Writing -- review \& editing.
\textbf{Pingquan Wang:} Resources, Writing -- review \& editing.
\textbf{Hailun Yuan:} Writing -- review \& editing.
\textbf{Hui Zhang:} Resources, Writing -- review \& editing.
\textbf{Xiaopeng Zhou:} Conceptualization, Methodology, Supervision, Project administration, Funding acquisition, Writing -- review \& editing.

\section*{Declaration of competing interest}
The authors declare that they have no known competing financial interests or personal relationships that could have appeared to influence the work reported in this paper.

\section*{Data availability}
Data will be made available on reasonable request.

\section*{Acknowledgements}
Supported by National Key R\&D grant from the Ministry of Science and Technology of China (Nos. 2023YFA1606200, 2023YFA1606201), grants from Beihang University "Ganwei" project (No. JKF-2025083066927).

\section*{Declaration of Generative AI and AI-assisted technologies in the manuscript preparation process}

During the preparation of this work the authors used ChatGPT and DeepSeek in order to improve the readability and language of the manuscript. After using these tools, the authors reviewed and edited the content as needed and take full responsibility for the content of the publication; the scientific content, all numerical results and all conclusions were produced by the authors. In addition, a generative AI image tool (Google Gemini) was used to produce the schematic drawing shown in Fig.~\ref{fig:combined-setup}(a); the authors compared every component, its arrangement and all labels with the actual experimental apparatus and take full responsibility for the accuracy of the drawing.

\bibliographystyle{elsarticle-num}
\bibliography{references}

\end{document}